\documentclass[10pt,twoside,BCOR7mm,DIV15,headinclude,footexclude,
               cleardoubleempty,idxtotoc]
{scrartcl}

\usepackage{natbib}
\usepackage[font=small,labelfont=bf]{caption}
\usepackage[english]{babel}
\usepackage{graphicx}
\usepackage{hyperref}
\usepackage{scrpage2}
\usepackage{ifthen}
\usepackage{booktabs}
\usepackage{amsmath}
\usepackage{amssymb}
\usepackage{multicol}
\usepackage{float}
\usepackage{hyperref}

\hypersetup{breaklinks=true
,colorlinks=true,linkcolor=black,urlcolor=blue
,citecolor=black}

\addto\captionsenglish{%
}

\makeatletter
\renewcommand{\@biblabel}[1]{}
\renewcommand{\@cite}[2]{%
{#1\ifthenelse{\boolean{@tempswa}}{,#2}{}}}
\makeatother
\ofoot{\thepage}
\ifoot{}

\setcapindent{0em}
\setheadsepline{1pt}

\setkomafont{pagehead}{\normalfont\sffamily}
\setkomafont{pagenumber}{\normalfont\rmfamily}

\makeatletter
\newcommand{\listofcontributions}{\@starttoc{con}}

\newcommand{\l@contribution} {\@dottedtocline{1}{1.5em}{2.3em}}
\makeatother

\newenvironment{contribution}{
\setcounter{section}{0}
\setcounter{figure}{0}
\setcounter{table}{0}
}{
\newpage
\lehead{}
\rohead{}
}

\begin{document}

\setlength{\baselineskip}{2.5ex}

\begin{contribution}
% EXAMPLE AND TEMPLATE FILE FOR PROCEEDINGS OF THE WOLF-RAYET WORKSHOP.
% PLEASE REPLACE THE TEMPLATE TEXT BY YOUR OWN ARTICLE.
% NOTE THAT YOU MUST NOT PROCESS THIS FILE, BUT THE MASTER FILE:
% latex masterfile; dvips masterfile
\newcommand{\msun}{\mbox{${M}_{\odot}$}}

% RUNNING AUTHOR: PUT AUTHOR NAMED HERE
\lehead{V.~V.\ Dwarkadas \& D.\ Rosenberg}

% RUNNING TITLE; SHORTEN THE TITLE IF NECESSARY
% IN CASE OF A ONE-PAGE CONTRIBUTION (POSTER),
% SQUEEZE AUTHORS AND TITLE IN THIS LINE (Author: Title ...)
\rohead{Wind Bubbles around Wolf-Rayet Stars}

\begin{center}
% FULL TITLE HEADING
{\LARGE \bf X-ray Emission from Ionized Wind-Bubbles around Wolf-Rayet
  Stars}\\
\medskip

% AUTHORS LIST
{\it\bf V.~V. Dwarkadas$^1$ \& D.\ Rosenberg$^2$}\\

% AFFILIATIONS
{\it $^1$Astronomy and Astrophysics, University of Chicago}\\
{\it $^2$Oak Ridge National Laboratory}

% ABSTRACT
\begin{abstract}
Using a code that employs a self-consistent method for computing the
effects of photoionization on circumstellar gas dynamics, we model the
formation of wind-driven nebulae around massive Wolf-Rayet (W-R)
stars. Our algorithm incorporates a simplified model of the
photo-ionization source, computes the fractional ionization of
hydrogen due to the photoionizing flux and recombination, and
determines self-consistently the energy balance due to ionization,
photo-heating and radiative cooling. We take into account changes in
stellar properties and mass-loss over the star's evolution. Our
multi-dimensional simulations clearly reveal the presence of strong
ionization front instabilities. Using various X-ray emission models,
and abundances consistent with those derived for W-R nebulae, we
compute the X-ray flux and spectra from our wind bubble models. We
show the evolution of the X-ray spectral features with time over the
evolution of the star, taking the absorption of the X-rays by the
ionized bubble into account. Our simulated X-ray spectra compare
reasonably well with observed spectra of Wolf-Rayet bubbles. They
suggest that X-ray nebulae around massive stars may not be easily
detectable, consistent with observations.
\end{abstract}
\end{center}

% TEXT OF THE PAPER, TWO-COLUMN STYLE
\begin{multicols}{2}

\section{Introduction:}
Massive stars ($>$ 8 M$_{\odot}$) lose mass throughout their lifetime,
via winds and eruptions, before ending their lives in a cataclysmic
supernova (SN) explosion. The interaction of this material with the
surrounding medium creates vast wind-blown cavities surrounded by a
dense shell, referred to as wind-blown bubbles. As the star evolves
through various stages, the mass-loss parameters will change,
affecting the structure of the bubble. When the star finally explodes,
the resulting SN shock wave will expand within the bubble, and the
dynamics and kinematics of the shock wave will depend on the bubble
parameters (\cite{vvd05,vvd07a,vvd07b}). Similarly, the relativistic
blast waves associated with gamma-ray bursts are expected to expand
within wind bubbles surrounding Wolf-Rayet (W-R) stars.  Using an
ionization-gasdynamics code, AVATAR, we compute the structure and
evolution of the wind-blown bubbles around massive stars.  Using the
ISIS package, we compute the X-ray spectrum from the simulations as
would be observed by the Chandra satellite, and compare to
observational data.

\section{The AVATAR Code:} 
We have further developed a code that combines photoionization from
the star with the gasdynamics, and used it to compute the structure of
wind bubbles around massive stars. The method operator splits the
contribution due to photoionization effects from the usual gas
dynamics, and utilizes a backward-Euler scheme together with a
Newton-Raphson iteration procedure for achieving a solution. The
effects of geometrical dilution and of column absorption of radiation
are considered. The gasdynamic algorithm makes use of a
multidimensional covariant implementation of well established Eulerian
finite difference algorithms (\cite{dr95}). A second-order (van Leer)
monotonic transport algorithm is used for advection of total mass and
the neutral component, and a third order piecewise parabolic scheme is
available. Tabulated functions are used to compute the collisional
ionization rate and cooling function. Shocks are treated using an
artificial viscosity. Grid expansion is available to study flow over
distance scales spanning several orders of magnitude. The algorithm
incorporates a simplified model of the photo-ionization source,
computes the fractional ionization of hydrogen due to the
photo-ionizing flux and recombination, and determines self-consistently
the energy balance due to ionization, photo-heating and radiative
cooling. In this, our method is superior to that of other
calculations, such as \cite{glm96,vlg05}, who use the on-the-spot
approximation and do not take the recombination time into account. It
is comparable to the work of \cite{ta11}.

\section{Evolution of the Wind-Blown bubble around a 40 M$_{\odot}$ star:} 
A model for a 40 $\msun$ star using the AVATAR code is shown in
Figure \ref{vikramfig:hydro}. Stellar parameters are adapted
from \cite{vlg05}. The evolution of the star can be divided into 3
main phases. An inhomogeneous pressure and density distribution
develops in the main-sequence (MS) stage, accompanied by vorticity
deposition near the inner shock. The inclusion of photo-ionization
results in the formation of a dense, lower temperature ($\sim 10^4$ K)
region of ionized material outside the wind bubble during the MS
phase. The nebula is fully ionized, and the ionization front is
trapped in the dense shell, which is unstable to ionization
instabilities. In the red supergiant (RSG) stage the surface
temperature of the star decreases considerably. Consequently the
ionizing radiation drops considerably and recombination reduces the
ionization fraction to $\sim$ 40\%, although this goes up again in the
W-R stage. The high-density RSG wind is followed by a higher momentum
W-R wind which breaks up the RSG shell, scattering the dense material
into a turbulent W-R nebula.

%-----------One-column figure -----------------------------------
% Note that only the [H] option is allowed for placing 1-column figures!
\begin{figure}[H]
\begin{center}
\includegraphics[width=\columnwidth]{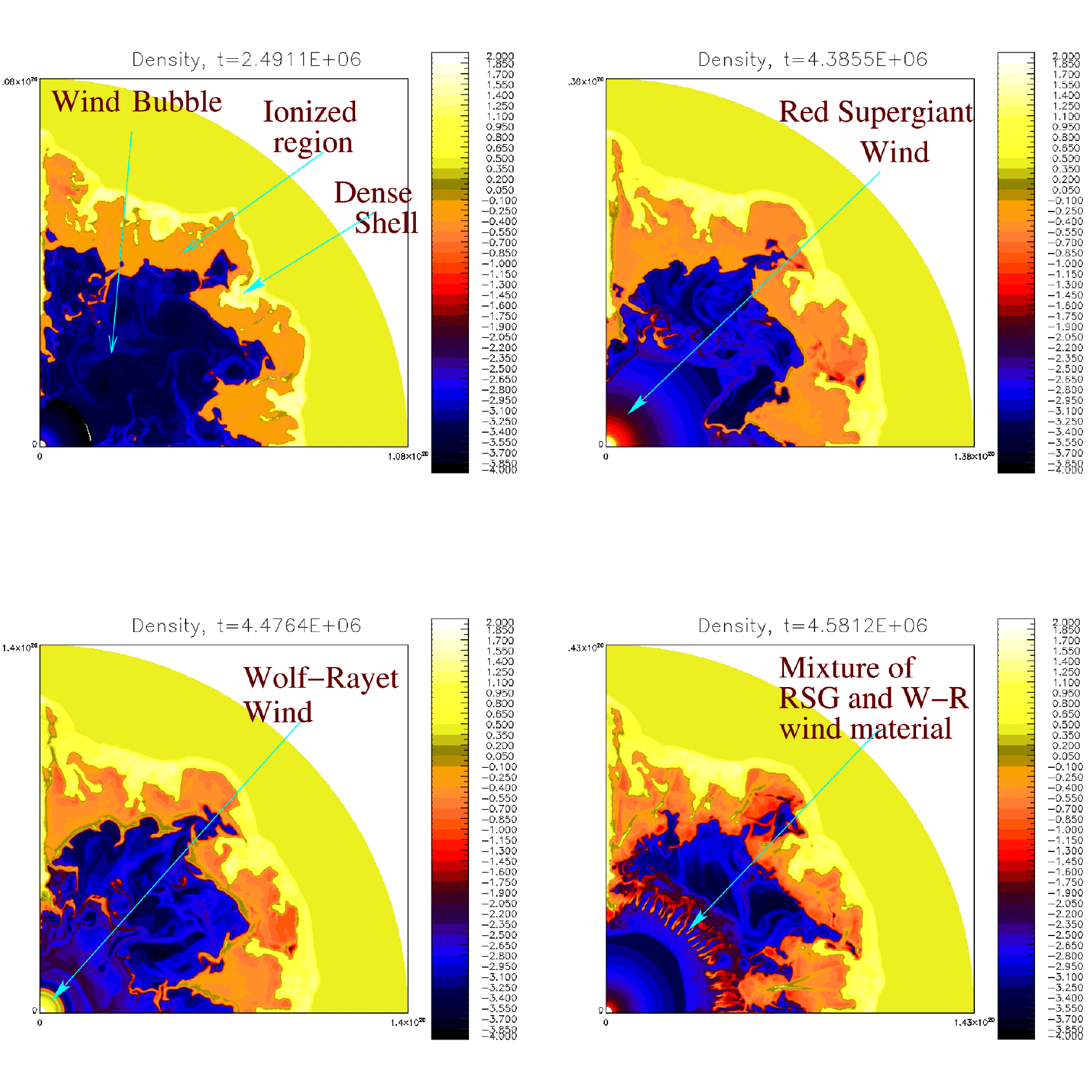}
\caption{Density snapshots from calculation of wind bubble evolution
  around 40 $\msun$ star (600 $\times$ 400 zones), computed with
  AVATAR with expanding grid (\cite{dr13}).  Time increases from left
  to right and top to bottom, and is listed at the top of each plot.
\label{vikramfig:hydro}}
\end{center}
\end{figure}
%-----------------------------------------------------------

We can use the simulations to illustrate the regions from which the
X-ray flux arises. We assume that thermal bremsstrahlung is the
dominant contributor to the X-ray flux, although line emission will
also play a role. The flux is then proportional to $n_e^2\,T^{1/2}$
where $n_e$ is the density and $T$ the temperature. We show maps of
this quantity at various times in Figure \ref{vikramfig:bremms}. The
normalization is arbitrary, however they are all normalized to the
same scale. Only zones having temperature $T > 10^6$ K are assumed to
emit X-rays. The densest clumps do not contribute because their
temperature is lower than 10$^6$ K. It is clear that the highest X-ray
flux arises in the W-R stage, and that the highest levels of X-ray
emission emanate from very small regions.

%-----------One-column figure -----------------------------------
% Note that only the [H] option is allowed for placing 1-column figures!
\begin{figure}[H]
\begin{center}
\includegraphics[width=\columnwidth]{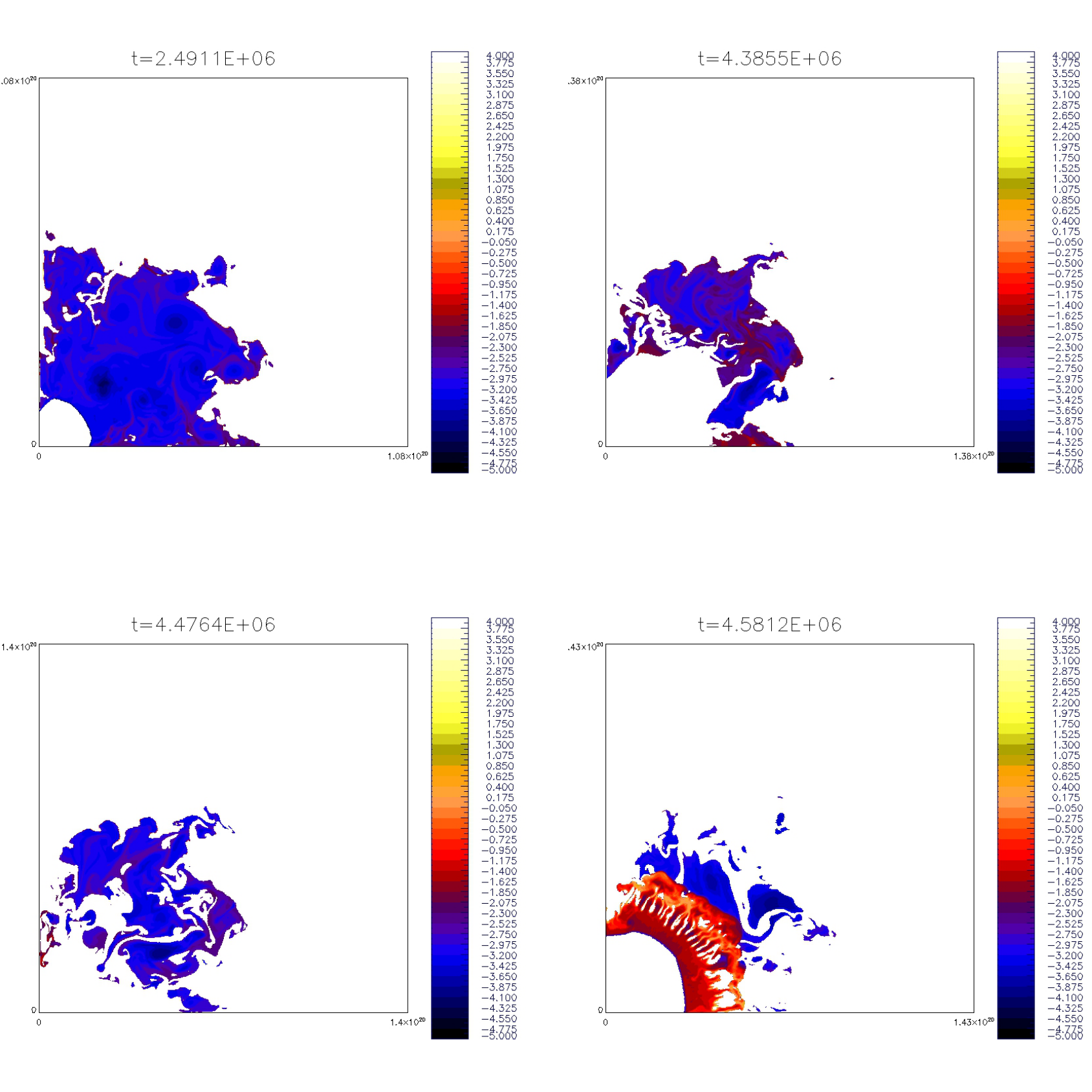}
\caption{Maps showing the area from which X-ray emission arises from
  the bubble in Figure \ref{vikramfig:hydro}. An increase of the X-ray flux
  in the W-R stage is seen.
\label{vikramfig:bremms}}
\end{center}
\end{figure}
%-----------------------------------------------------------

\section{THE X-RAY SPECTRUM OF THE BUBBLE:} 
In order to calculate the X-ray emission accurately, we use the ISIS
package (\cite{hd00}).  The data are read into ISIS, the spectrum
calculated for every zone, taking the absorption outside that zone
into account, and added together.  A point source is assumed.  We use
the VMEKAL model in XSPEC to model the spectrum, since it is the one
most commonly used to fit the observed X-ray spectra.  Herein we
present (Figure \ref{vikramfig:spectra}) X-ray spectra as would be detected
by the Chandra ACIS instrument for an object at 1.5kpc distance,
integrated over 50,000s.  Foreground absorption of 2.  $\times$
10$^{20}$ cm$^{-2}$ is added to absorption due to the neutral material
surrounding the bubble, which varies, and can be as high as
2. $\times$ 10$^{21}$ cm-2 (total NH in units of 10$^{22}$ cm$^{-2}$).
Line broadening is based on the underlying fluid velocity.  Solar
abundances (\cite{ag89}) are used for the MS and RSG
stages. Abundances in the W-R phase are from \cite{chuetal03} for the
W-R bubble S308.

%----------- Double-column figure -----------------------------------
\begin{figure*}[htb]
\begin{center}
\includegraphics
  [width=\textwidth]{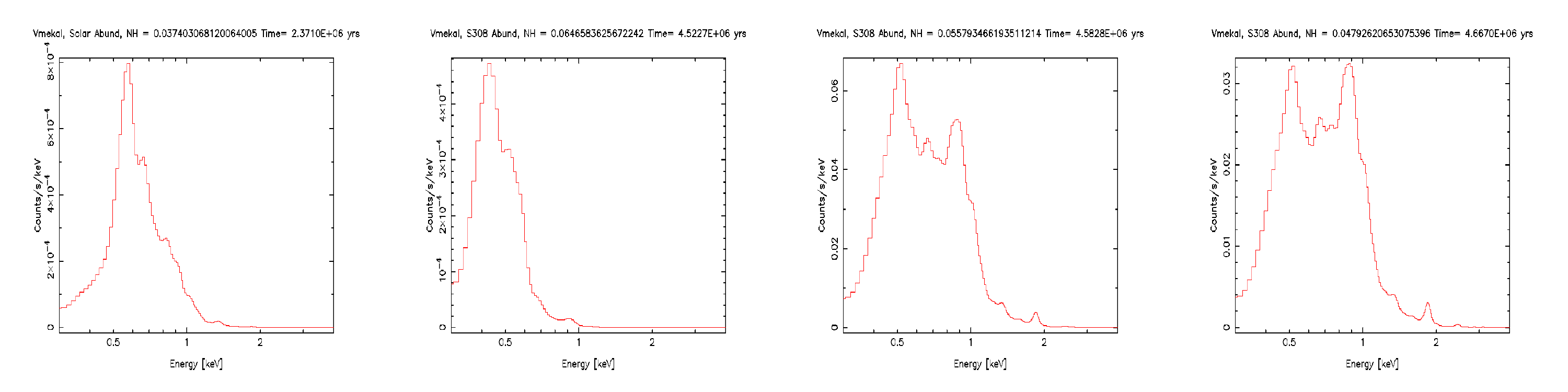}
\caption{ Simulated X-ray spectra for a point source calculated from
  the simulation shown in Figure \ref{vikramfig:hydro}. The leftmost
  image is from the MS phase, the next three are from the W-R
  phase. Note that as time increases in the W-R stage, the intensity
  of the spectrum (counts/sec/keV) also increases.
\label{vikramfig:spectra}}
\end{center}
\end{figure*}
%--------------------------------------------------------------------

Our spectra are comparable in shape and temperature to observed
spectra (Figure \ref{vikramfig:speccomp}). Our simulations show that
the X-ray emission in the MS phase is generally too weak to be
detected by current X-ray satellites.  It is higher in the W-R stage,
but still difficult to detect.  The X-ray emission depends strongly on
the density within the bubble, which is a function of the density of
the material around the nebula. Thus it is possible that a higher
ambient density leads to a higher intensity, although unless it was
all ionized the absorption would correspondingly increase. The general
low intensity is in agreement with observations, which have only
detected 3 W-R nebulae, and no MS nebulae around massive stars
(\cite{chuetal06,tggc14,tgcg15})
%----------- Double-column figure -----------------------------------
\begin{figure*}[htb]
\begin{center}
\includegraphics
  [width=\textwidth]{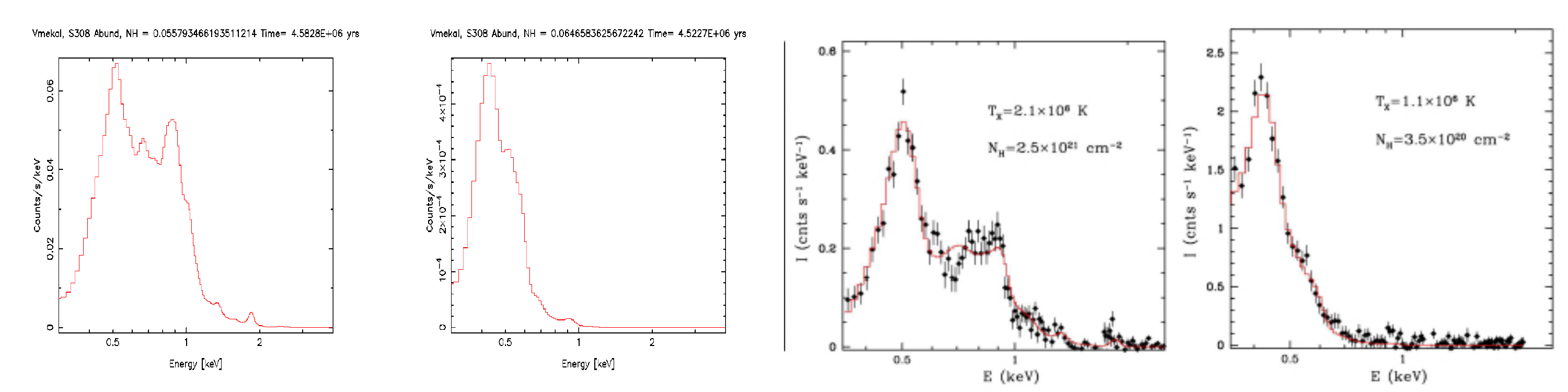}
\caption{Simulated and Observed spectra.  Panel 1 resembles observed
  Chandra spectrum for NGC 6888 (panel 3) and Panel 2 XMM spectrum of
  S308 (panel 4), from \cite{chuetal06}. Although the spectral shapes
  are similar, the counts are significantly lower in our simulated
  spectra as compared to the data. This seems to be generally true
  over all our simulations, in agreement with the fact that diffuse
  X-ray emission is not generally observed.
\label{vikramfig:speccomp}}
\end{center}
\end{figure*}
%--------------------------------------------------------------------

{\bf Acknowledgements:} Support for VVD is provided by Chandra grants
TM9-0001X and TM5-16001X issued by the CXO, which is operated by SAO
for and on behalf of NASA under contract NAS8-03060. DR acknowledges
resources of the OLCF at ORNL, which is supported by the Office of
Science of the U.S. DOE under Contract No. DE-AC05-00OR22725.

\bibliographystyle{aa} % style aa.bst
\bibliography{myarticle}

\end{multicols}

\end{contribution}

%%-------------------------------------------------------

\end{document}